\documentclass[fleqn,usenatbib]{mnras}

\usepackage{newtxtext,newtxmath}
\usepackage{gensymb}

\usepackage[T1]{fontenc}
\usepackage{soul}
\DeclareRobustCommand{\VAN}[3]{#2}
\let\VANthebibliography\thebibliography
\def\thebibliography{\DeclareRobustCommand{\VAN}[3]{##3}\VANthebibliography}

\usepackage{graphicx}	
\usepackage{amsmath}	
\RequirePackage{fix-cm}
\usepackage{xcolor} 
\title[Intensity Interferometry with 0.25~m Telescopes]{Intensity Interferometer Results on Sirius with 0.25~m Telescopes}

\author[T. J. Mozdzen et al.]{Thomas J. Mozdzen,$^{1}$\thanks{E-mail: tmozdzen@asu.edu }
Richard M. Scott,$^{1}$
Ricardo R. Rodriguez,$^{2}$
and Philip D. Mauskopf$^{1}$
\\
$^{1}$Arizona State University, School of Earth and Space Exploration, Tempe, Arizona 85287\\
$^{2}$Case Western Reserve University, Cleveland, Ohio 44106\\
}

\date{Accepted XXX. Received YYY; in original form ZZZ}

\pubyear{2025}

\begin{document}
\label{firstpage}
\pagerange{\pageref{firstpage}--\pageref{lastpage}}
\maketitle


\begin{abstract}
We present the successful measurement of the squared visibility of Sirius at a telescope separation of 3.3~m using small 0.25~m Newtonian-style telescopes in an urban backyard setting. The primary science goal for small-scale intensity interferometers has been to measure the angular diameters of stars. Recent advances in low jitter time-tagging equipment and Single Photon Avalanche Detectors have made the detection of second-order photon correlation signals feasible with small low-cost telescopes. Using Sirius as a target star, we observe a photon count rate of $\sim$1.9~Mcps per detector with matched 1.2~nm wide filters at 589.3~nm and measured the spatial squared visibility at a telescope separation of 3.3~m to be $|V_{12}(3.3\text{m})|^2~=~0.94\pm0.16$. The measured detection significance is $\sim7 \sigma$ after 13.55~h of integration.The uncertainty in the measured visibility includes uncertainty in the instrument response function.The squared visibility agrees closely with the expected value of $0.94\pm0.01$. These results demonstrate that using small low-cost telescopes is feasible for intensity interferometry of bright stars. This enables a simple scaling in sensitivity by further realistic improvements in the instrument response jitter as well as increasing both the number of spectral bands and the number of telescopes towards systems capable of resolving objects such as quasars, white dwarfs, and galactic Cepheid variable stars.
\end{abstract}



\section{Introduction}
One of the key frontiers in observational astronomy is improving high angular resolution to obtain better images and measure the physical sizes of astronomical objects. One approach to achieving this is by building larger single-aperture telescopes, both in space (e.g., JWST) and on the ground, where adaptive optics are used to overcome atmospheric refraction. The next generation of ground-based telescopes, such as GMT, TMT, and ELT, will feature aperture sizes over 20~m, with angular resolution down to 10~milliarcseconds.

Another path to higher resolution involves interferometric techniques using optical and near-infrared interferometers like GRAVITY at VLTI, MROI, NPOI, and CHARA \cite{vlti_status, mroi_status, chara_status, npoi_status}. These instruments combine light from multiple telescopes across different baselines, detecting interference patterns in real time. Recent technical advances have revived amplitude interferometry performance, with more improvements on the horizon \cite{Eisenhauer}.

At longer wavelengths, in the radio and millimeter-wave bands, digital interferometry records the time-varying electric field at remote locations and combines the signals in software. This Very Long Baseline Interferometry (VLBI) technique allows for extremely large baselines from telescopes distributed across Earth, resulting in high angular resolution. For instance, millimeter-wave VLBI enabled the Event Horizon Telescope (EHT) to capture images of supermassive black holes with 25~microarcsecond resolution \cite{Akiyama_2019, EHTCollaboration_2022}.

Digital beam combination is also possible at optical wavelengths using intensity correlations, first detected from stars in the 1950s by \cite{BROWN1956}. A major advantage of intensity interferometry is that it can combine signals digitally, similar to VLBI, and is insensitive to atmospheric turbulence. Over the past decade, advances in photon detection, timestamp resolution, and more affordable equipment have renewed interest in intensity interferometry. Current detectors and time taggers are capable of high efficiency single photon detection with timing resolution of 10s of picoseconds or better approaching the limit from atmospheric path variations \cite{Dalal2024}. Intensity interferometry can complement amplitude interferometry measurements on objects such as radial oscillations of Cepheids, rotational flattening of fast-rotating stars, and gravitational-wave-emitting binaries \cite{MAGIC2024}.

The smallest system to measure intensity correlations from a star so far used a 0.5~m aperture telescope with a beamsplitter for a zero-baseline configuration on Sirius \cite{Karl2024}. \cite{Horch_2022} constructed an interferometer with two portable 0.6~m telescopes to study stars such as Altair, Deneb, and Vega. \cite{Matthews2022, Matthews2023} combined a portable 1~m telescope with a fixed 1.54~m telescope to study the H$\alpha$ accretion disk around the Be star $\gamma$ Cas and used two 1.8~m auxiliary telescopes at the VLTI for further observations. VERITAS \cite{2020Veritas, Acharyya_2024} used four 12~m Cherenkov telescopes to collect data on multiple baselines, while MAGIC \cite{MAGIC2024} utilized two 17~m Cherenkov telescopes, and HESS \cite{HESS2023} used four 12~m telescopes. The original HBT experiments were performed with two 6.5~m telescopes at Narrabri \cite{HBT1956}.

In pursuit of demonstrating a low-cost arrayed intensity interferometry system, we show that small, portable, low-cost telescopes can successfully detect a correlation signal and \textcolor{black}{accurately measure the visibility}. Our system uses two 0.25 m telescopes, Single Photon Avalanche Detectors (SPADs), a time-to-digital converter, and custom autoguiding optics to measure intensity correlations from Sirius. The SPADs and timestamping equipment used in this system provide a 12-fold improvement in combined jitter (147~ps~vs.~1800~ps~FWHM) compared to our previous system \cite{Adrian2016, 2017MNRASGena}. 

The following sections discuss the theory behind intensity correlations, particularly the relationship between signal-to-noise ratio (SNR), coherence time ($\tau_{\text{coh}}$), \textcolor{black}{and squared visibility}; describe the interferometer components; and present the results from field observations of Sirius ($\alpha$ CMa). We conclude with a description of methods to expand the system, such as arraying telescopes and using more frequency bands \cite{LaiOlivier2018Cote} to reduce collection time and enable observations of fainter sources.

\section{Intensity Interferometry Theory}
\label{sec:Theory}

In this section, we derive equations relating contrast degradation, coherence time ($\tau_{\text{coh}}$), visibility, and the significance of detection SNR for binned data.

\subsection{Intensity Correlation Fluctuations and Baseline Dependence}

The normalized temporal correlation of the intensities from a common source at two measurement points, $I_1$ and $I_2$, is:
\begin{equation}    
 \frac{\langle I_1(t)I_2(t+\tau) \rangle}{\langle I_1 \rangle\langle I_2\rangle}~=~g^{(2)}(\tau)~=~1+~  |V_{12}(d)|^2|\gamma_{11}(\tau)|^2,
 \label{eq:CorrelationEquation}
\end{equation}
where $d$ is the projected baseline distance between the two detectors, $\tau$ is the time delay, $\gamma_{11}(\tau) = g^{(1)}(\tau)$ is the temporal autocorrelation function of the radiation, and $|V_{12}(d)|^2$ is the squared visibility \cite{HBT_Book_1974,MANDEL1963, Foellmi2009}, which for a uniform disk star is described by the squared Airy function. The shape of $|V_{12}(d)|^2$ depends on the observed wavelength of light, $\lambda$, the angular size of the source, $\theta$, and the separation between the two observations, $d$. 

\subsection{Contrast Degradation and Coherence Time}
The Arizona State University Stellar Intensity Interferometer (ASUSII) as well as most other intensity interferometry systems satisfy $\tau_{\text{res}}~\gg~\tau_{\text{coh}}~\approx~1/\Delta\nu ~\sim$~1~ps, which results in contrast degradation. For a non-polarized chaotic light source like thermal radiation, the intensity fluctuations in the different polarizations and spatial modes are uncorrelated and the normalized amplitude of $|g^{(1)}(\tau)|$ is further reduced by a factor of $\sqrt{n_\text{M}}$ where $n_\text{M}$ is the number of spatial and/or polarization modes. The intrinsic coherence time of the light source is given by,
\begin{equation}
    \tau_{\text{coh-i}}=\int_{-\infty}^{+\infty} |g^{(1)}(\tau)|^2 d\tau,
    \label{eq:tcoh}
\end{equation}
and the measured intensity correlations will be spread out over an effective time, 
\begin{equation}
    \tau_{\text{res}}=\int_{-\infty}^{+\infty} |M_{11}(\tau)|^2 d\tau,
    \label{eq:tres}
\end{equation}
where $|M_{11}(\tau)|^2$ is the measurement response function which we model as a Gaussian with a variance, $\sigma_{\text{res}} = \tau_{\text{res}}/\sqrt{2 \pi}$. The measured \textcolor{black}{coincidence counts distribution, $h_{\rm meas}(\tau)$, is the convolution} of the intrinsic coherence function and the measurement function resulting in:

\begin{equation}
    \textcolor{black}{h_{\rm meas}(\tau)} = 1+~ \frac{1}{n_\text{M} \tau_{\text{res}}} |V_{12}(d)|^2\left[|g^{(1)}(\tau)|^2 \circledast |M_{11}(\tau)|^2 \right].
    \label{eq:meas}
\end{equation}

Because the integral of the convolution of two functions is equal to the product of the integrals of the separate functions, the integral of the second term in Eq.~\ref{eq:meas} is given by (for $n_\text{M} = 2$):
\begin{equation}
    \int_{-\infty}^{\infty}\frac{d\tau}{2 \tau_{\text{res}}} |V_{12}(d)|^2\left[|g^{(1)}(\tau)|^2 \circledast |M_{11}(\tau)|^2 \right] = \frac{\tau_{\text{coh-i}}}{2}|V_{12}(d)|^2,
\end{equation}

\textcolor{black}{We recognize the quantity in the integrand as half of the measured coherence time at a projected baseline $d$, $\tau_{\text{coh-m}}$, giving}
 \begin{equation}
     |V_{12}(d)|^2=\frac{\tau_{\text{coh-m}}}{\tau_{\text{coh-i}}}.
     \label{eq:IntegratedSignal_simple}
 \end{equation}
 
In Appendix~\ref{sec:SNR}, we derive the SNR equation 
\begin{equation}
        SNR=\frac{\tau_{\text{coh-i}}}{2} \left(\frac {R_1 R_2 T_{\text{int}}}{\sqrt{2}~\tau_{\text{res}}}\right)^{1/2}|V_{12}(d)|^2,
        \label{eq:SNR}
\end{equation}
\textcolor{black}{which is based upon count rates $R_1~\text{and}~R_2$, integration time $T_\text{int}$, system resolution $\tau_\text{res}$, intrinsic coherence time $\tau_\text{coh-i}$, and squared visibility $|V_{12}(d)|^2$. In Appendix~\ref{sec:SNR_data_section} we derive an $SNR_\text{data}$ metric based solely on histogram bin counts.}

\section{System Equipment}
\begin{figure} 
    \centering
    \includegraphics[width=0.7\linewidth]{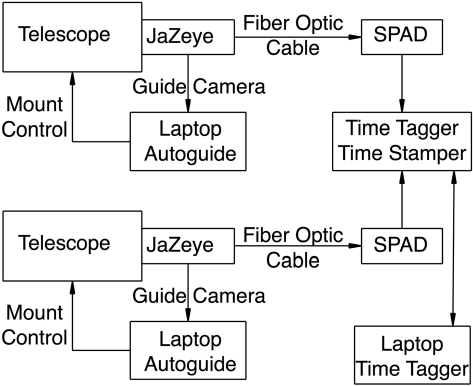}
    \caption{Block diagram of the two telescope intensity interferometer.}
    \label{fig:BlockDiagramField}
\end{figure}

The interferometer block diagram is shown in Fig.~\ref{fig:BlockDiagramField}. Each telescope has an ASU designed JaZeye optical system attached to it that filters and couples the light from the telescope to fiber optic cables. The fiber cables feed the light to the SPADs. The pulses from the SPADs are sent to the time-to-digital converter (Time Tagger) that time stamps the incoming photons that are then stored on a hard drive for later analysis. Computers are used to run the Time Tagger, monitor the field/guide cameras on the JaZeyes, and run the autoguider system for the telescopes. The following subsections describe each of the main interferometer components.

\subsection{Telescopes}
\begin{figure} 
    \centering
    \includegraphics[width=1\linewidth]{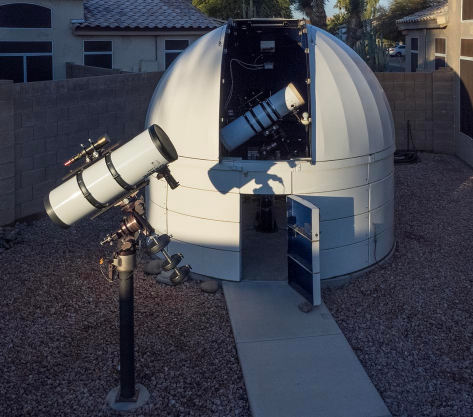} 
    \caption{The two telescopes set up at the JaZ Observatory.}
    \label{fig:InterferometerTelescopes}
\end{figure}

Two small telescopes (0.25~m diameter) are used at the JaZ Observatory for initial testing of the interferometer. One telescope is an f/4.6 Lurie-Houghton (LH) that was custom designed and built by one of the authors. The other telescope is a commercial Meade f/4 Schmidt-Newtonian (SN). Both telescopes are on German equatorial mounts for tracking the star under measurement and have on-axis \mbox{autoguiders} to keep the star centered on the fiber cable. Fig.~\ref{fig:InterferometerTelescopes} shows the telescopes at the JaZ observatory set up on the shortest baseline.

\subsection{JaZeye Optical System Description}
\begin{figure} 
    \centering
    \includegraphics[width=0.95\linewidth]{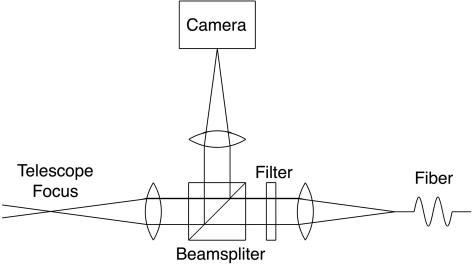} 
    \caption{Schematic of the JaZeye showing its components and light path.}
    \label{fig:OpticalPath}
\end{figure}

The JaZeye is the optical system that distributes incoming light from the target star to both the guide camera and the SPAD. A fiber cable is used to present a lighter load on the JaZeye than having the SPAD connected directly to it. The JaZeye is a compact assembly that provides narrowband filtering and includes a camera for on-axis autoguiding (see section \ref{sec:guiding}). A schematic of the JaZeye optical path is shown in Fig.~\ref{fig:OpticalPath}.

\begin{figure} 
    \centering
    \includegraphics[width=1\linewidth]{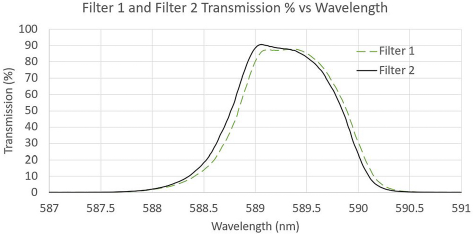} 
    \caption{Transmission Profiles of Sodium Filters 1 and 2. The expected intrinsic coherence time is 0.78~ps.}
    \label{fig:FilterProfile}
\end{figure}

Narrowband filtering is necessary to increase the coherence time, which increases the contrast of the correlation and reduces the average count rate from the star to below the maximum count rate of the SPADs. The filter is specified to be 1.2~nm wide centered at 589.45~nm with a peak transmission of 97 per cent and out-of-band transmission less than 0.004 per cent. We verified the filter performance by measuring the transmission with a Cary 5000 spectrophotometer. Fig.~\ref{fig:FilterProfile} shows the measured transmission of the two filters overlaid to show the nearly identical transmission characteristics. The peak transmission is 88 per cent, centered at 589.25~nm with an overlapping transmission width of 1.15~nm. We transform the filter response into a power density spectrum and, using Eq.~\ref{eq:tcoh} and Eq.~\ref{eq:gamma11}, the expected intrinsic coherence time is 0.78~ps. To achieve the full performance of the filter, the light from the telescope is collimated with an aspheric lens. After filtering, the light is focused on the fiber cable using an identical lens to maintain the image scale of the telescope. The lenses are broadband anti-reflection coated to minimize reflections and maximize transmission.

\begin{figure} 
    \centering
   \includegraphics[width=1\linewidth]{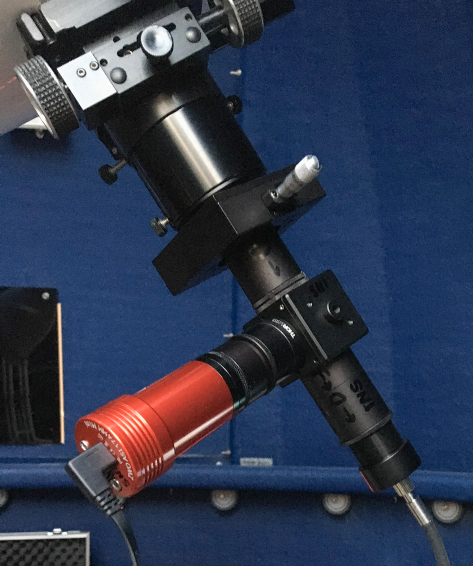} 
   \caption{A JaZeye mounted on one of the telescopes.}
    \label{fig:JaZeye}
\end{figure}

A non-polarizing beamsplitter is inserted into the collimated light beam prior to the filter to pick off a small fraction (10\%) of the incoming light for the camera used to monitor the star in real-time and as input to the autoguiding system. The camera aids in the initial positioning and focusing of the star on the fiber cable and enables autoguiding during data collection. One of the JaZeyes mounted on a telescope is shown in Fig.~\ref{fig:JaZeye}.

\subsection{Detectors, Time Tagger, and Fiber Optic Cables}

This section describes the ASUSII photon counting system, consisting of two SPADs, a Time Tagger, and two fiber optic cables.

\subsubsection{Single Photon Detectors}
The detectors are Micro Photon Devices (MPD) PDM \$PD-100-CTD-FC Single Photon Avalanche Detectors (SPAD). They have the following specifications: active area diameter~=~100~$\mu$m; photon efficiency at the ASUSII filter's wavelength (589.25~nm)~=~40 per cent (with FC fiber connector); dark count rate~=~50 cps; dead time~=~77~ns; and pulse width~=~17~ns. The factory jitter test data for our specific SPADs is 32~ps. The dark count rate is significantly lower than the typical photon count rate of 1 to 2~Mcps, so it has virtually no effect on the results.

\label{sec:TimeTagger}

\subsubsection{Time Tagger}
Time stamps of the detected photons are recorded using the Time Tagger Ultra from Swabian Instruments which has single-channel 1$\sigma$ jitter of 42~ps. The combined jitter of  the SPADs and Time Tagger is 147 ps FWHM. The second-order correlation signal, g$^{(2)}(\tau)$, is detected by creating a histogram of the differences in arrival times of pairs of photons, which arrive in different detectors, after adjusting for differences in travel times. When the light is coming from a star, the difference in the distance to each telescope is constantly changing (see section \ref{sec:OPD}). This uneven time travel is canceled by periodically adding a time offset to one of the channel's timestamps.

\subsubsection{Fiber Optic Cable}
\label{sec:fiber_dispersion}
The fiber cables that were used for the present observations are 3~m long, 105~um step-index multimode fibers with a 0.22~NA and armored for ruggedness. The next observing runs will be performed with shorter, 1~m, fiber cable to reduce its temporal dispersion contribution by two-thirds (see section~\ref{sec:Uncertainty}).

\subsection{Telescope Guiding}
\label{sec:guiding}
\begin{figure}
    \centering
    \includegraphics[width=0.95\linewidth]{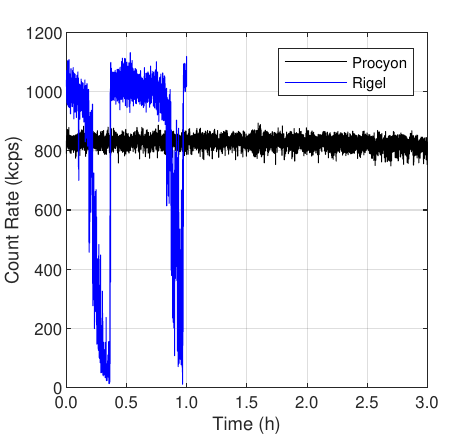} 
    \caption{ The benefit of on-axis guiding vs off-axis guiding is shown to provide steady count rates for several hours in contrast to 20 to 30 minute stability.}
    \label{fig:Guiding}
\end{figure}

Early observing runs showed that we needed to improve the telescope's ability to track the star accurately for maximum photon collecting efficiency. The initial observing runs used separately mounted guide scopes, and the count rate kept dropping off every 20 to 30 minutes, forcing us to manually re-center the star to boost the count rate. The count rate dropping over time with the original guiding system compared to the on-axis guiding system we developed is shown in Fig.~\ref{fig:Guiding}.
Testing showed that the star was drifting off the fiber over a short time period.

The camera on the JaZeye is now used as the input to the commercial PHD2 autoguiding software running on a laptop computer. Once calibrated, PHD2 sends control pulses to the telescope mount to keep the star centered on the same location of the guide camera sensor. This modification to the interferometer allows accurate tracking of the star for the entire observing period.

\section{Observations and Measurements} 
\label{sec:Observations and Measurements}

\label{sec:On-Sky Observations}
Data collection on Sirius was performed over three nights in January and February of 2024 in Chandler, Arizona, accumulating 13.55 h of observations, using two 0.25~m telescopes separated by 3.3~m. Table~\ref{tab:observations-table-2024} shows the starting dates and times, observing duration, count rates, and temperature changes. Sirius was chosen as the target star because it met two key criteria: 1) potentially high count rates (2.0~Mcps) at the wavelength of the narrowband filters and the detector's quantum efficiency; and 2) an angular diameter that allowed for a strong correlation signal at the 3.3~m spacing, yet was not too small, ensuring that the roll-off in signal strength could be observed at larger baselines in a residential backyard setting.

\begin{table}
\begin{center}
\caption{\label{tab:observations-table-2024}Observations performed on Sirius at the JaZ Observatory in 2024.}
\begin{tabular}{|c|c|c|c|c|c|c|} 
 \hline
 Date &   Start &Mins& Time & CH1  & CH2&Temp\\ [0.5ex] 
  (2024)     &Time&Before&   on Sky    &Rate &Rate& Begin -\\ 
      &  (MST)    &Transit &(h) &(Mcps) & (Mcps) &End (\degree C)\\ 
 \hline
 Jan 30 &21:13  &84.2 &4.25 & 1.9 & 2.0&16->11 \\ 
 Feb 13 & 20:34 & 68.2&4.88 & 1.7 & 1.9&14->7 \\
 Feb 14 & 20:30 & 68.3&4.41 & 1.8 & 2.0&14->7 \\
\\
 Totals  & & &13.55 & 1.8 & 2.0 \\
 \hline
\end{tabular}
\label{tab:Observations2024Sirius}
\end{center}
\end{table}

\subsection{Expected Count Rates and System Efficiency}
Using (\cite{Rybicki})
\begin{equation}
    F=\pi B \left(\frac{R}{r}\right)^2  \left(\frac{1}{h\nu_0}\right)~\Delta\lambda~A_\text{T}, (\text{photons/s})
    \label{eq:photonRate}
\end{equation}
where $B$ is the Planck blackbody radiation law, $R$ is the radius of Sirius, $A_\text{T}$ is the collecting area of the telescope, and $r$ is the distance to Sirius. With the ASUSII telescopes and a 1.2~nm wide filter, we estimated an expected count rate per telescope of $15\times10^6$ cps before taking losses into consideration. Table \ref{tab:LightLoss} lists the major photon losses in the equipment. Beyond this 81 per cent loss, atmospheric absorption at the JaZ Observatory (362~m elevation) adds another 10 per cent, reducing the expected count rate to 2.6 Mcps. With a photons arriving every 390 ns, the 80~ns SPAD dead time causes an additional 20~per cent loss. The final estimated count rate of 2.04~Mcps aligns with observed rates of 1.7 to 2.0~Mcps.


\begin{table}
    \begin{center}
    \caption{Transmission efficiency of system components taking into account optical losses and SPAD efficiency. The total optical efficiency portion is 49 per cent.}
    \begin{tabular}{lcc}
    \hline
              & Lurie-Houghton & Meade SN10 \\
    Component & Telescope & Telescope \\ [0.4ex]
    & (per cent) & (per cent) \\
    \hline
    
    \textbf{Total Telescope Efficiency} & 71 & 72 \\ 
    \textbf{Total JaZeye Efficiency} & 74 & 74 \\
    \textbf{Total Fiber Efficiency} & 92 & 92 \\
    \textbf{Detector Efficiency @ 589~nm} & 39 & 39 \\
    \\
    \textbf{Total System Efficiency} & 19 & 19 \\
    \hline
    \end{tabular}
    \label{tab:LightLoss}
    \end{center}
\end{table}

\subsection{Optical Path Length Differences and Baselines}
\label{sec:OPD}
The optical path length difference (OPD) is given by \cite{UVWplane}

\begin{equation}
    \label{eq:OPD_eq}
    \begin{split}
    OPD  =   &~~~~B_N(-\text{sin}(lat)\text{cos}(h)\text{cos}(dec) + \text{cos}(lat)\text{sin}(dec)) \\
            &- B_E(\text{sin}(h)\text{cos}(dec)) \\
            &+ Z_{UP}(\text{cos}(lat)\text{cos}(h)\text{cos}(dec) + \text{sin}(lat)\text{sin}(dec)),
    \end{split}
\end{equation}
where $B_N, B_E,$ and $Z_{UP}$ are the local components of the baseline distance vector $\boldsymbol{B}$; $lat$ is the latitude of the telescopes, $h$ is the hour angle, and $dec$ is the declination of the target star.

A 50~m coax cable with a measured delay of 201~ns was inserted in one of the Time Tagger inputs in order to move correlations away from $\tau~=~0$ on the Time Tagger for possible cross-talk issues. For the current measurement, our goal was to obtain a strong detection of the correlation signal. The maximum projected baseline was 3.3~m, and the expected squared visibility was expected to be $\geq$~93.7 per cent of its maximum value, 100 per cent. In future measurements with a longer baseline, we will take into account the projected baseline as the target moves across the sky.

\subsection{Data Processing}
Each night, data comprising timestamps and telescope channels is stored and processed the next day with a custom Matlab program. The program reads 64 million timestamps at a time, adjusting the timestamps of one channel to compensate for the optical path delay difference between the telescopes. With the observed count rates, 64 million events are collected every 15~s, with delay adjustments typically ranging from 10 to 15~ps per data chunk. Histograms from multiple nights are then combined using a bin size of 1~ps. We create additional histograms with larger bin sizes to assess bin size effects on the correlation signal, as overly large bins can smear the signal, while very small bins can reduce smoothness.

\subsection{Results on Sirius}
\label{sec:Sirius_Results}

Combining three nights of data we detected a g$^{(2)}$ correlation signal with a squared visibility of $|V_{12}(3.3~\text{m})|^2 = 0.94\pm0.16$. Table~\ref{tab:SiriusObservationStats} shows the measured noise compared to the theoretical Poisson noise ($\sigma_{\text{STD}}$ vs. $\sigma_{\text{Poisson}}$) in the histograms of the correlated photon counts for different bin sizes. The measured signal to noise ratio, SNR$_{\text{data}}$ = \textcolor{black}{H$_\text{Fit}\sqrt{N_{\text{eff}}}/\sigma_{\text{STD}}$}, is calculated from the total integrated signal divided by the noise where the signal is spread over an effective number of bins given by $N_{\text{eff}} = \tau_{\text{res}}/(\sqrt{2}\tau_{\text{bin}})$ assuming a Gaussian instrument response function with the best fit value for $\tau_{\text{res}}$  (see Appendix~\ref{sec:SNR} and ~\ref{sec:SNR_data_section}).  

\textcolor{black}{Detection significance values ranged between 6.6 and 7.1 $\sigma$ using Eq.~\ref{eq:SNR_data} and these values are also consistent with Eq.~\ref{eq:SNR}.}  \textcolor{black}{The detection significance, SNR$_{\text{data}}$, quantifies how much the measured data stands out compared to what would be expected from random variations alone, accounting for both the size of the signal and the level of fluctuations in the data.} \textcolor{black}{This metric is important for small telescopes as integration times must exceed one hour to begin to detect a signal above the noise level. Also, the SNR formulae are useful for predicting performance of future systems.}

\begin{figure}
    \centering
    \includegraphics[width=0.95\linewidth]{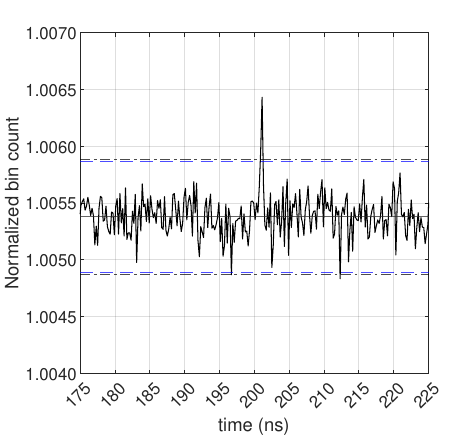} 
    \caption{Normalised g$^{(2)}$ correlation histogram from Sirius using two telescopes from three runs with bin size of 200~ps. The dashed lines are normalised $\pm$3 Std-RMS and $\pm$3~Poisson-Noise. The un-normalised statistics are shown in Table~\ref{tab:SiriusObservationStats}}.
    \label{fig:SiriusCorrelation}
\end{figure}

Fig.~\ref{fig:SiriusCorrelation} displays the 200~ps bin histogram normalised by the theoretical Poisson noise calculated from the total counts per bin over time differences from 175~ns to 225~ns, where the dashed horizontal lines mark the normalised $\pm3~\sigma_\text{STD}$ and $\pm~3~\sigma_\text{Poisson}$ noise limits. The normalisation slightly exceeds 1.0 due to a reduction in the average count rates over time (slightly violating the stationary random process requirement) between detectors due to increasing atmospheric extinction as Sirius’s altitude lowers. The choice of bin size is a trade-off between including sufficient data points for fitting and preserving the correlation signal shape (see Table~\ref{tab:SiriusObservationStats}).

\begin{figure}
    \centering
    \includegraphics[width=0.95\linewidth]{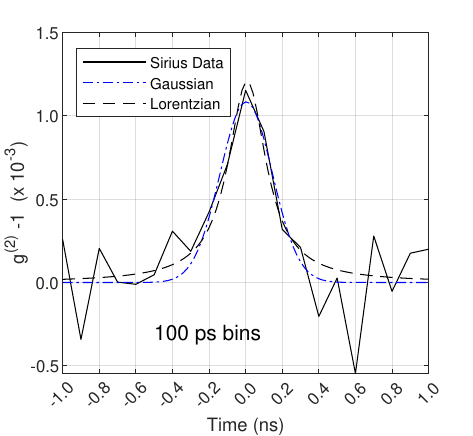} 
    \caption{Fitted Normalized Gaussian and Lorentzian curves for the Sirius $g^{(2)}$~-~1 correlation signal using 100 ps bins. The Gaussian FWHM is 325~ps and the peak value is $1.07\times10^{-3}$.}
    \label{fig:SiriusNormalizedCorrelationFit}
\end{figure}

Both Gaussian and Lorentzian curves were fit to the normalized 100~ps bin histogram (Fig.\ref{fig:SiriusNormalizedCorrelationFit}).
Simultaneously fitting the amplitude ($H_{\text{fit}}$) and width ($\sigma_{\text{fit}}$) of a Gaussian to the normalized data gave $H_{\text{fit}} = 1.072 \pm 0.177 \times 10^{-3}$ with $\sigma_{\text{fit}}~=~136.4 \pm 26$~ps, resulting in $\tau_{\text{res}}~\equiv~\sigma_{\text{fit}}\sqrt{2\pi}~=~342\pm65$~ps (Table~\ref{tab:FittingStats}). The coherence time of the correlation signal can be calculated from twice the area under the curve (Eq.~\ref{eq:IntegratedSignal_simple}), $\tau_{\text{coh-m}}~=~2 H_{\text{fit}} \tau_{\text{res}}~=~0.733 \pm 0.121$~ps, where the uncertainty in the coherence time was calculated using the covariance matrix of the two-component fit. We found that the uncertainty in the amplitude was anti-correlated with the uncertainty in the width of the instrument response function. From Eq.~\ref{eq:IntegratedSignal_simple}, $|V_{12}(3.3~\text{m})|^2~=~\tau_{\text{coh-m}}/\tau_{\text{coh-i}}~=~0.94\pm0.16$, which agrees with the expected squared visibility of $0.94\pm0.01$ based on Sirius's known diameter and the projected baseline averaged over the hour angles of operation.

\begin{table}
\caption{These values are derived from the histogram binned data. The measured detection significance \textcolor{black}{SNR$_\text{data}$} was calculated using Eq.~\ref{eq:SNR_data}. \textcolor{black}{H$_\text{data}$ is the value of the signal bin with the highest count value minus the mean of the bins away from the peak, and  H$_\text{fit}$ is the peak height of the Gaussian fit to the bin data.}}
\begin{center}
\begin{tabular}{|l|c|c|c|c|} 
 \hline
 Statistic &  25 ps & 50 ps & 100 ps & 200 ps \\ [0.5ex] 
     &   bin  & bin    & bin & bin\\ 
 \hline
  Mean~($10^7$)                            & 0.443 & 0.887 & 1.77   & 3.55 \\ 
  $\sigma_{\text{STD}}$                    & 2132  & 3016  & 4182   & 5760 \\
  $\sigma_{\text{Poisson}}$                & 2100  & 2970  & 4200   & 5940 \\
  H$_{\text{Data}}$                        & 5900  & 11172 & 20007  & 35301\\
  H$_{\text{Data}}$/$\sigma_{\text{STD}}$  & 2.77  & 3.70  & 4.78   & 6.13 \\
  H$_{\text{Fit}}$                         & 4657  & 9492  & 19010  & 36174\\
  H$_{\text{Fit}}/\sigma_{\text{STD}}$     & 2.18  & 3.14  & 4.55   & 6.28 \\
  $\tau_{\text{res}}$~(ps)                 & 325   & 316   & 342    & 363  \\
  N$_{\text{eff}}^{\text{bins}}$           & 9.20  & 4.47  & 2.44   & 1.28 \\
  SNR$_{\text{data}}$                      & 6.63  & 6.65  & 7.10   & 7.11 \\
 \hline
\end{tabular}
\label{tab:SiriusObservationStats}
\end{center}
\end{table}

\subsection{Sources of Uncertainties}
\label{sec:Uncertainty}
The FWHM value for the Gaussian fit on Sirius was $321\pm61$~ps, which is greater than the 147~ps FWHM jitter value for the Time Tagger and the SPADs. Additional sources of jitter in the measured data include: the temperature dependence of the 50~m delay cable, uncalibrated optical path delays between the two telescopes, and temporal dispersion in the fiber cables.
Models for temporal dispersion in the 3~m long 100~$\mu$m diameter multimode fiber cables give a maximum time difference of 168~ps \cite{STEPNIAK2017}. This can be reduced by using shorter cables while still minimizing mechanical strain on the JaZeyes. The measured change in the delay cable using a temperature chamber was approximately 1~ns per 100~\degree C. We estimated that the temperature of the delay cable fell by 5~\degree C to 7~\degree C over the Sirius observing runs. The variation in the delay of the cable is estimated to be 50 to 70~ps.

\begin{table}
\caption{Gaussian fitting results on the measured correlation signal of Sirius. The area, height, and sigma of the Gaussian fits are listed in the table.} 
\begin{center}
\begin{tabular}{c|c|c|c|c|c|} 
 \hline
 $\tau_\text{{coh-m}}$  &  Fit Height  & Sigma & $\tau_{\text{res}}$ & $|V_\text{12}(3.3~\text{m})|^2$ \\
              & (Contrast  & Fit & derived &\\
      & Factor)&&\\
  (ps) & $\times10^{-3}$ & (ps) & (ps)& \\
 \hline
0.73~$\pm$0.12& 1.07~$\pm$0.18 & 136~$\pm26$ &342~$\pm65$&0.94~$\pm$0.16\\
 \hline
\end{tabular}

\label{tab:FittingStats}
\end{center}
\end{table}

\section{Conclusions}
The availability of relatively low-cost single photon detectors, time taggers, simple optical interfaces, and affordable small telescopes has made it possible to develop a low-cost intensity interferometer system. Using two small backyard (0.25~m) telescopes tracking Sirius and collecting data for 13.55~hours with an on-axis guiding system to maximize the photon collection efficiency, we measured the squared visibility at a maximum baseline separation of $3.3$~m to be $|V_{12}(3.3~\text{m})|^2 = 0.94\pm0.16$, which agrees with the expected value, 0.94$\pm$0.01 based upon the
known visibility function of Sirius. The detection significance is $\sim 7 \sigma$.  The effective resolution time, $\tau_{\text{res}} = 342$~ps was partly limited by the combined Time Tagger and SPAD jitter of 147~ps FWHM, and by fiber cable dispersion jitter. With near future enhancements, a system jitter of 60~ps is feasible which will decrease collecting times for a given SNR by a factor of up to 6. Compared to the original HBT measurements, the ASUSII collected over 10,000 times fewer photons (smaller collecting area and narrower bandwidth) for the same SNR.

Building on the demonstration that a pair of 0.25 m telescopes can detect the second-order coherence signal from a bright star, we can scale this system in order to observe fainter objects, such as an 8th-magnitude white dwarf, using baselines of 500 to 1000 meters with existing clock synchronization technology, such as the White Rabbit system (\cite{WhiteRabbit}). With the existing system, the photon count rates and corresponding SNR would be $\sim$6000 times lower which could be compensated by using two 5x4 arrays of 0.76~m telescopes such as the LFAST arrays (\cite{lfast}) (200x more photons), 20~ps system jitter (4x SNR boost), and 100 spectral channels (10x SNR boost) to achieve the same SNR as the current system achieved on Sirius in 13.5~h. \cite{Dalal2024} provides a detailed analysis of scaling parameters, highlighting their trade-offs, and recommends that initial scaling efforts prioritize characterization of bright sources. With ongoing technological advancements and growing research activity in the field, the future looks promising for intensity interferometers to once again make significant contributions to observational astronomy.

\section*{Acknowledgments}
The Breakthrough Starshot project provided funding for work in single photon detection. The Quantum Collaborative, led by Arizona State University, provided valuable expertise and resources for this research project. The Quantum Collaborative connects top scientific programs, initiatives, and facilities with prominent industry partners to advance the science and engineering of quantum information science. We thank Todd Hodges for his work on this project as a graduate student. The work he accomplished for his thesis "Measurement, Detection, and Parameter Estimation of Single Photon Correlations" \cite{Hodges2022} set the stage for the recent follow-on work with the new SPADs and Time Tagger. Todd also assisted us with early backyard field work.
We graciously thank Swabian Instruments, OptoElectronic Components, and Micro Photon Devices for loaning us time taggers and SPADs for evaluation prior to purchase and for excellent post-purchase support. We thank our reviewer

for his careful reading of the manuscript and valuable feedback to improve its quality.

\section*{Data Availability}
The data underlying this article will be shared on reasonable request to the corresponding author.



\bibliographystyle{mnras}
\bibliography{02_references} 



\appendix
\section{Relationship between coherence time and bandwidth}
\label{sec:Spectral Shapes}
We can calculate the intrinsic coherence time of the light source, $\tau_{\text{coh-i}}$, from the shape of the power spectral density of the incoming light. Because the autocorrelation function is the Fourier transform of the frequency spectrum:
\begin{equation}
    g^{(1)}(\tau) = FT(G(\nu))/ \int_{0}^{\infty}G(\nu)d\nu,
    \label{eq:gamma11}
\end{equation}
where $G(\nu)$ is the spectral density of the light common to both band-pass filters. Eq. \ref{eq:tcoh} then gives us $\tau_{\text{coh-i}}$. If no analytic form is available, the calculation can be done numerically. For the analytic Top Hat, Gaussian, and Lorentzian spectral shapes, the intrinsic coherence time and the FWHM of the spectrum are related by:
\begin{align}
\Delta \nu_{\text{FWHM}}~&=~ \frac{1}{\tau_{\text{coh-i}}}~&=~\frac{1.00}{\tau_{\text{coh-i}}}~&~\text{Top Hat}\\
\Delta \nu_{\text{FWHM}}~&=~ \left(\frac{2 \text{ln}(2)}{\pi}\right)^{1/2}\frac{1}{\tau_{\text{coh-i}}}~&=~\frac{0.66}{\tau_{\text{coh-i}}}~&~\text{Gaussian} \\
\Delta \nu_{\text{FWHM}}~&=~ \frac{1}{\pi \tau_{\text{coh-i}}}~&=~\frac{0.32}{\tau_{\text{coh-i}}}~&~\text{Lorentzian}
\end{align}

\section{Measured Intensity Correlations and Signal to Noise Ratio}
\textcolor{black}{We used the SNR equations to form estimates of the integration time required to obtain a significant detection sigma based on count rates, system resolution, coherence time of the filters, and the squared visibility. These equations are also useful as a metric to compare and evaluate future intensity interferometry system configurations.}
\subsection{Count Rate based derivation}
\label{sec:SNR}
Following \cite{Purcell1956,MANDEL1963,HBT_Book_1974}, we calculate the signal from single photon counting detectors from the intensity fluctuations due to bunching and random Poisson statistics, where we bin the pairs of photons by their temporal separation. The number of coincident counts as a function of time delay, $N_{\text{hist}}(\tau) = N_{\text{Poisson}}(\tau) + N_{\text{Signal}}(\tau)$, in a correlation histogram for unpolarized light ($n_\text{M} = 2$) from a star with telescopes separated by a distance, $d$ and histogram bin size, $\delta \tau_{\text{bin}}$, is given by:
\begin{equation}
    N_{\text{hist}}(\tau_{\text{bin}}) = R_1R_2 \delta \tau_{\text{bin}}T_{\text{int}}\left[1+\frac{\tau_{\text{coh-i}}}{2\tau_{\text{res}}}|M_{11}(\tau_{\text{bin}})|^2|V_{12}(d)|^2\right],
    \label{eq:CoincidenceCount}   
\end{equation}

where $\tau_{\text{coh-i}}$ is the intrinsic coherence time, $R_1$~and~ $R_2$ are the individual photon count rates, $T_{\text{int}}$ is the integration time, and $\tau_{\text{bin}}$ takes on integer multiples of $\delta \tau_{\text{bin}}$ values. Eq. \ref{eq:CoincidenceCount} is the unnormalized version of Eq. \ref{eq:meas} with the normalization factor, $N_{\text{Poisson}}(\tau_{\text{bin}})= R_1 R_2 \delta \tau_{\text{bin}} T_{\text{int}}$.

The optimum SNR is obtained by a sum over bins weighted by the instrument response function, $|M_{11}(\tau)|^2$ which is equivalent to multiplying the signal amplitude by an effective number of bins,  $N_{\text{eff}} = \tau_{\text{res}}/(\alpha \delta \tau_{\text{bin}})$ where $\alpha = 1$ for a top hat instrument response (see Appendix~\ref{sec:Spectral Shapes}). For a Gaussian instrument response function, $\alpha = \sqrt{2}$, and the number of signal coincident counts integrated over the effective number of bins is given by: 
\begin{equation}
    \begin{split}
   N_{\text{signal}}^{\text{tot}} &=N_{\text{eff}} R_1R_2 \delta \tau_{\text{bin}}T_{\text{int}}\frac{\tau_{\text{coh-i}}}{2\tau_{\text{res}}}|V_{12}(d)|^2 \\
   &= R_1R_2  T_{\text{int}}\frac{\tau_{\text{coh-i}}}{2\sqrt{2}}|V_{12}(d)|^2,
    \end{split}\label{eq:TotSignal}
\end{equation}
and the number of random coincidences for unpolarized light, $N_{\text{Poisson}}$, is given by 
\begin{equation}
   N_{\text{Poisson}}^{\text{tot}}=N_{\text{eff}} R_1 R_2 \delta \tau_{\text{bin}} T_{\text{int}} = R_1 R_2 \tau_{\text{res}} T_{\text{int}}/\sqrt{2}. 
   \label{eq:PoissonNoise}
\end{equation}

The SNR of the integrated signal, 
\begin{equation}
SNR = N_{\text{signal}}^{\text{tot}}/\left(N_{\text{Poisson}}^{\text{tot}}\right)^{1/2},
\end{equation}
leads to
\begin{equation}
        SNR=\frac{\tau_{\text{coh-i}}}{2} \left(\frac {R_1 R_2 T_{\text{int}}}{\sqrt{2} \tau_{\text{res}}}\right)^{1/2}|V_{12}(d)|^2.
        \label{eq:SNRappendix}
\end{equation}

\subsection{Histogram count based derivation}
\label{sec:SNR_data_section}
\textcolor{black}{
We can also use the measured count values in the histogram bins to determine the SNR significance of detection metric. Using the $N_{\text{eff}}$ concept, the total signal count in the bins, is given by
}
\begin{equation}
\textcolor{black}{
    N_\text{signal} = H_{\text{fit}}N_{\text{eff}},}
\end{equation}
\textcolor{black}{and the total standard deviation count is determined by adding the standard deviation, $\sigma_\text{STD}$, of each of the $N_{\text{eff}}$ bins in quadrature, giving}
\begin{equation}
    \textcolor{black}{N_\text{STD} = {\sigma_\text{STD}}\sqrt{N_{\text{eff}}}}
\end{equation}
\textcolor{black}{resulting in}
\begin{equation}
   \textcolor{black}{ SNR_\text{data} = H_{\text{fit}}\sqrt{N_{\text{eff}}}/{\sigma_\text{STD}}}
    \label{eq:SNR_data}
\end{equation}



\bsp	
\label{lastpage}
\end{document}